! program = pdflatex
\documentclass[12pt]{article}
\usepackage[hmargin=2.54cm,vmargin=2.54cm]{geometry} 
\usepackage{geometry}
\usepackage{amsmath,amssymb,setspace,fancyhdr,geometry,url,color}
\usepackage{subfigure,graphicx,caption}
\usepackage{lscape,amsthm}
\usepackage[authoryear,round]{natbib}
\RequirePackage{lineno}
\allowdisplaybreaks[1]
\allowbreak
\title{A composite likelihood approach to computer model calibration
  using high-dimensional spatial data}
\author{Won Chang, Murali Haran, Roman Olson, and Klaus Keller}
 
\newtheorem{proposition}{Proposition}
\begin{document}
\linenumbers
\newcommand{\E}{\ensuremath{\mbox{E}}}
\newcommand{\Cov}{\ensuremath{\mbox{Cov}}}
\newcommand{\bA}{\ensuremath{\mathbf{A}}}
\newcommand{\barB}{\ensuremath{\mathbf{\bar{B}}}}
\newcommand{\bolda}{\ensuremath{\mathbf{a}}}
\newcommand{\cl}{\ensuremath{c\ell}}
\newcommand{\bomega}{\ensuremath{\boldsymbol{\omega}}}
\newcommand{\bM}{\ensuremath{\mathbf{M}}}
\newcommand{\bK}{\ensuremath{\mathbf{K}}}
\newcommand{\bG}{\ensuremath{\mathbf{G}}}
\newcommand{\bJ}{\ensuremath{\mathbf{J}}}
\newcommand{\bIn}{\ensuremath{\mathbf{I}_n}}
\newcommand{\bIN}{\ensuremath{\mathbf{I}_N}}
\newcommand{\Tau}{\ensuremath{\mbox{T}}}
\newcommand{\Eta}{\ensuremath{\mbox{H}}}
\newcommand{\Sigom}{\Sigma_{\boldsymbol{\omega}}}
\newcommand{\bmu}{\ensuremath{\mathbf{\mu}}}
\newcommand{\bnu}{\boldsymbol{\nu}}
\newcommand{\sigep}{\sigma_{\boldsymbol{\epsilon}}^2}
\newcommand{\bZ}{\ensuremath{\mathbf{Z}}}
\newcommand{\barZ}{\ensuremath{\mathbf{\bar{Z}}}}
\newcommand{\bz}{\ensuremath{\mathbf{z}}}
\newcommand{\bS}{\ensuremath{\mathbf{S}}}
\newcommand{\bX}{\ensuremath{\mathbf{X}}}
\newcommand{\bR}{\ensuremath{\mathbf{R}}}
\newcommand{\bB}{\ensuremath{\mathbf{B}}}
\newcommand{\bW}{\ensuremath{\mathbf{W}}}
\newcommand{\bw}{\ensuremath{\mathbf{w}}}
\newcommand{\ba}{\ensuremath{\mathbf{a}}}
\newcommand{\bb}{\ensuremath{\mathbf{b}}}
\newcommand{\bC}{\ensuremath{\mathbf{C}}}
\newcommand{\bD}{\ensuremath{\mathbf{D}}}
\newcommand{\bQ}{\ensuremath{\mathbf{Q}}}
\newcommand{\bP}{\ensuremath{\mathbf{P}}}
\newcommand{\bY}{\ensuremath{\mathbf{Y}}}
\newcommand{\barY}{\ensuremath{\bar{\mathbf{Y}}}}
\newcommand{\be}{\ensuremath{\mathbf{e}}}
\newcommand{\bs}{\ensuremath{\mathbf{s}}}
\newcommand{\bu}{\ensuremath{\mathbf{u}}}
\newcommand{\bv}{\ensuremath{\mathbf{v}}}
\newcommand{\bphi}{\ensuremath{\boldsymbol{\phi}}}
\newcommand{\btheta}{\ensuremath{\boldsymbol{\theta}}}
\newcommand{\bLambda}{\ensuremath{\boldsymbol{\Lambda}}}
\newcommand{\bepsilon}{\ensuremath{\boldsymbol{\epsilon}}}
\newcommand{\bdelta}{\ensuremath{\boldsymbol{\delta}}}
\newcommand{\boldeta}{\ensuremath{\boldsymbol{\eta}}}
\newcommand{\balpha}{\ensuremath{\boldsymbol{\alpha}}}
\newcommand{\bpsi}{\ensuremath{\boldsymbol{\psi}}}
\newcommand{\bxi}{\ensuremath{\boldsymbol{\xi}}}
\newcommand{\bzero}{\ensuremath{\mathbf{0}}}
\newcommand{\bk}{\ensuremath{\mathbf{k}}}
\maketitle
\doublespacing
\begin{abstract}
Computer models are used to model complex processes in various disciplines. 
Often, a key source of uncertainty in the behavior of complex computer models is uncertainty due to unknown model input parameters. Statistical computer model calibration is the process of inferring model parameter values, along with associated uncertainties, from observations of the physical process and from model outputs at various parameter settings. Observations and model outputs are often in the form of high-dimensional spatial fields, especially in the environmental sciences. Sound statistical inference may be computationally challenging in such situations. Here we introduce a composite likelihood-based approach to perform computer model calibration with high-dimensional spatial data. While composite likelihood has been studied extensively in the context of spatial statistics, computer model calibration using composite likelihood poses several new challenges. We propose a computationally efficient approach for Bayesian computer model calibration using composite likelihood. We also develop a methodology based on asymptotic theory for adjusting the composite likelihood posterior distribution so that it accurately represents posterior uncertainties. We study the application of our new approach in the context of calibration for a climate model. 
\end{abstract}
\section{Introduction} \label{section:introduction}
{ Complex computer models are often used to approximate real-world processes. These models enable us to conduct virtual experiments that are useful for studying and understanding complex physical phenomena. A central source of uncertainty regarding computer models, and hence the behavior of the process they are approximating, stems from uncertainty about the value of model input parameters. 
It is, however, often possible to learn about model parameter values from observations of the system being modeled. Computer model calibration, the methods used to learn about these parameters, involves finding model parameter settings that produce computer model outputs that are most compatible with the observed realization of the process. Statistical computer model calibration is a formal approach to parameter inference based on observations and on computer model output at various parameter settings. A sound approach to computer model calibration accounts for various sources of uncertainties such as measurement error and model structural errors, and results in a probability distribution that summarizes our knowledge about the parameters. Quantifying uncertainties about the parameters carefully is important as this allows for a rigorous quantification of uncertainties about projections based on the model. Here we consider computer model calibration for problems where the observations and the model output are in the form of spatial data. 
 
Computer model calibration can pose nontrivial inferential challenges. In many applications computer model runs are computationally expensive. In this case, model runs are often available at only a limited number of parameter settings. A popular method to overcome this hurdle is the Gaussian process approach
\citep[cf.][]{sacks1989design, kennedy2001bayesian}. This method enables calibration with a limited number of model runs using probabilistic interpolation between the model runs.
However, this approach faces computational challenges when applied to computer model output that are in the form of high-dimensional spatial data, which are increasingly common in modern science and engineering applications \cite[see e.g.][]{higdon2009posterior,bhat2010computer,bhat:hara:tonk:kell:2009,chang2013fast}.
 
Some approaches have been developed recently to resolve these computational issues  \citep[e.g.][]{bayarri2007computer,Higdon2008,bhat:hara:tonk:kell:2009,chang2013fast}. In this manuscript we propose a new Bayesian approach for calibration with high-dimensional spatial data using composite likelihood methods.
 
The basic idea of composite likelihood   \citep{besag1975statistical,besag1977efficiency,lindsay1988composite} is to approximate the original likelihood
as a product of computationally cheaper likelihoods. This approach can be easily adapted
for spatial modeling in various ways such as conditional likelihood \citep{vecchia1988estimation, stein2004approximating}, pairwise likelihood \citep{heagerty1998composite, curriero1999composite, ribatet2009bayesian}, and block likelihood \citep{caragea2006approximate, eidsvik2011estimation}.
 Here we construct a calibration method based on block composite
likelihood. In particular, we adopt the idea of hybrid composite
likelihood proposed by \cite{caragea2006approximate} that relies on two components: (i)  dependence between block means and (ii) dependence within each block conditioning on its block mean.  This composite likelihood approach allows for a substantial reduction in the computational burden for maximum likelihood inference with high-dimensional spatial data. Also, this opens up possibilities for flexible spatial covariance structure that vary depending on each
block. Moreover, since the composite likelihood from the block composite likelihood framework  is a valid probability model, no further justification is necessary for
its use in Bayesian inference. 
 
The remainder of this paper is organized as follows. In Section 2 we
outline the basic model calibration framework using Gaussian random
fields. In Section 3 we formulate the Bayesian calibration model using block composite likelihood, discuss relevant asymptotic theory and explain how Godambe information may be used to adjust posterior uncertainty when using composite likelihood. In Section 4 we describe an application of our method to a climate model calibration problem using 2-dimensional spatial patterns of ocean temperature change and a relevant simulated
example. Finally, in Section 5, we conclude with a discussion and future directions for research. 
 
\section{Calibration using Gaussian Processes} \label{section:CalibrationGP}
Here we introduce our computer model calibration framework which consists
of two stages: model emulation and parameter calibration
\citep{bayarri2007computer, bhat:hara:tonk:kell:2009,
  chang2013fast}. We first construct an `emulator',
which is a statistical model interpolating the computer model outputs as well as
providing interpolation uncertainties \citep{sacks1989design}. Using
the emulator, we find the posterior density of computer model
parameters while taking into account important sources of
uncertainty including interpolation uncertainty, model-observation
discrepancy, and observational error \citep{kennedy2001bayesian}.

We will use the following notation henceforth.  $Y(\bs,\btheta)$ is the computer
model output at the spatial location $\bs \in \mathcal{S}$ and the
parameter setting $\btheta \in \boldsymbol{\Theta}$.\ $\mathcal{S}$ is
the spatial field that we are interested in, usually a subset of
$\mathbb{R}^2$ or $\mathbb{R}^3$.\ $\boldsymbol{\Theta}\subset \mathbb{R}^q$ is the open
set of all possible computer model parameter settings with an integer $q \ge 1$. 
 Let $\left\{\btheta_1,\dots,\btheta_p\right\}\subset \boldsymbol{\Theta} $ be a collection of $p$ design points in the parameter space and $\left\{\bs_1,\dots,\bs_n\right\}\subset \mathcal{S}$ be the set of $n$ model grid locations.\
$\bY_i=(Y(\bs_1,\btheta_i),\dots,Y(\bs_n,\btheta_i))^T$ is computer model output at the model grid locations at the parameter setting $\btheta_i$.\  The concatenated $np\times 1$
vector of all computer model outputs is $\bY=(\bY_1^T,\dots,\bY_p^T)^T$. Note that typically $p \ll n$ since computer model runs with
high-resolution are computationally expensive. Finally, we let $Z(\bs)$ be an observation at  spatial location $\bs$ and $\bZ=(Z(\bs_1),\dots,Z(\bs_n))^T$ be the observational data, a spatial process observed at $n$ locations. \\
{\bf \noindent Model Emulation Using Gaussian Proccesses.} Following \cite{bhat:hara:tonk:kell:2009} and \cite{chang2013fast}, we
construct a Gaussian process that interpolates computer model outputs
as follows
\begin{equation*}
\bY \sim N(\mathbf{X}\boldsymbol{\beta},\Sigma(\bxi_y)),
\end{equation*}
where $\bX$ is an $np\times b$ covariate matrix containing all the
spatial locations and climate parameters (that is, $\bs_1,\dots,\bs_n$ and
$\btheta_1,\dots,\btheta_p$) used to define the covariance matrix
$\Sigma(\bxi_y)$.\ $\boldsymbol{\beta}$ and $\bxi_y$ are the vectors of regression coefficients
and covariance parameters respectively. We construct an interpolation process by
finding the maximum likelihood estimate (MLE) of these
parameters. This interpolation model provides the predictive distribution of a
computer model run at any given location $\bs \in \mathcal{S}$ and
$\btheta \in \boldsymbol{\Theta}$ \citep{sacks1989design}. We call
this predictive process an emulator and denote it by $\eta(\bs,\btheta)$.
Note that, throughout this paper, $\boldsymbol{\beta}$ is set to be $\mathbf{0}$ since the Gaussian process provides enough flexibility in modeling the output process.
 
{\bf \noindent Model Calibration Using Gaussian Random Processes.} We model the observational data $\bZ$ by the following model,
\begin{equation} \label{eqn:calibrationmodel}
\bZ=\boldeta(\btheta^*)+\bdelta,
\end{equation}
where $\btheta^*$ is the true or fitted value of computer model parameter for the observational data \citep{bayarri2007computer}, $\boldeta(\btheta^*)=\left(\eta(\bs_1,\btheta^*),\dots,\eta(\bs_n,\btheta^*) \right)^T$ is the emulator output at $\btheta^*$ on the model grid,  and $\bdelta=\left(\delta(\bs_1),\dots,\delta(\bs_n) \right)^T$ is a term that includes both data-model discrepancy as well as observational error. The discrepancy process $\delta(\bs)$ is also modeled as a Gaussian process with spatial covariance between the locations $\bs_1,\dots,\bs_n$. Model calibration with high-dimensional spatial data leads to computational challenges as described in the following section.

\section{Calibration with High-Dimensional Spatial Data} \label{section:our method}
 
In this section we briefly examine the challenges in model calibration using
high-dimensional spatial data and the existing approaches to the problem. We
then proceed to the formulation of our composite likelihood approach.
\subsection{Challenges with High-Dimensional Spatial Data}
The basic challenge with the approach in Section \ref{section:CalibrationGP} stems from the fact
that the computational cost for a single likelihood evaluation is $\mathcal{O}(n^3p^3)$. For large $n$,  evaluating the
likelihood function repeatedly when using algorithms like Markov chain Monte Carlo (MCMC) can become computationally prohibitive. One can reduce the computational cost by assuming a separable covariance structure between the spatial dependence and the dependence due to computer model parameters, but the computational cost is still $\mathcal{O}(n^3)$, and hence does not scale well with $n$.
The current
approaches to overcome such limitation for high-dimensional data rely
on dimension reduction or basis expansion. The dimension reduction
approaches \citep{bayarri2007computer,chang2013fast} map the original
output into a lower dimension and exploit the uncorrelated nature of
the low-dimensional processes to speed up the computation. The basis
expansion approaches \citep{bhat:hara:tonk:kell:2009, Higdon2008} use
a basis representation of model output that results in a reformulated
likelihood with a lower computational cost. 
Here we introduce a somewhat different approach that relies on the block composite likelihood for spatial data \citep{caragea2006approximate,eidsvik2011estimation}.

\subsection{Composite Likelihood for Model Calibration\label{section:our method}}
In this framework, we partition the spatial field $\mathcal{S}$ into small blocks
to avoid the computational issues related to high-dimensional
data. In Section \ref{section:application} we describe an example of how such a partition may be constructed in practice.
The block composite likelihood method substitutes the original likelihood by a composite
likelihood that utilizes the spatial blocks, thereby resulting in a likelihood function that requires much less computational effort.
In particular, we adopt the block composite likelihood formulation by \cite{caragea2006approximate}. This framework assumes
conditional independence between outcomes in different blocks
given the block means, and the dependence between blocks is modeled through the
covariance between block means. Note that this framework gives a valid
probability model, and therefore the posterior distribution defined
using the composite likelihood function based on this approach is also a valid
probability model. Obtaining a valid probability model is important because we are embedding the likelihood within a Bayesian approach; having a valid probability model automatically assures us that the resulting posterior distribution is proper when all the prior distributions used are proper.

We divide the spatial area for the computer model
output into $M$ different blocks and denote the output for each block
by $\bY_{(1)},\dots,\bY_{(M)}$. Note that the blocks are made according to the spatial field, not the parameter space, because the number of computer model runs is usually quite limited due to the high computational costs of running the model. However, in principle our approach may be extended to blocking in parameter space as well if the number of model runs is also large. Let $n_{i}$ denote the number of computer model outcomes in the $i$th block. We denote the spatial locations in the $i$th block by $\bs_{i1},\dots,\bs_{in_i}$. Each $\bY_{(i)}$ is a stack of $(n_i-1)$-dimensional spatial output for $p$ different parameter settings;
\begin{equation*}
\bY_{(i)}= \left( Y(\bs_{i1},\cdot)^T,Y(\bs_{i2},\cdot)^T,\dots,Y(\bs_{i n_i-1},\cdot)^T \right)^T,
\end{equation*}
where $Y(\bs_{ij},\cdot)=\left(Y(\bs_{i j},\btheta_1),\dots,Y(\bs_{i j},\btheta_p) \right)^T$ is the $p\times1$ vector of computer model
outcomes for all the parameter settings $\btheta_1,\dots,\btheta_p$. Note that we omit one spatial location for each block in defining the output
vectors to avoid degeneracy. We let
$\barY_{(i)}=\frac{1}{n_i}\sum_{j=1}^{n_i} \left(Y(\bs_{i j},\btheta_1),\dots,Y(\bs_{i j},\btheta_p)\right)^T$ be the
$p$-dimensional mean vector of model outcomes for the $i$th block. That is, means for the spatial block consisting of same set of locations across all model parameter settings. We define the vector of all block means by $\barY=\left(\barY_{(1)}^T,\dots,\barY_{(M)}^T\right)^T$. Similarly, we divide the observational data into $M$ blocks in the same way
and omit one observation for each block to have $\bZ_{(1)},\dots,\bZ_{(M)}$, the vectors of observational data in different blocks. We let $\barZ_{(i)}=\frac{1}{n_i}\sum_{j=1}^{n_i}Z(\bs_{ij})$ be the $i$th
block mean of observational data and $\barZ=\left(\barZ_{(1)},\dots,\barZ_{(M)}\right)^T$ be the collection of them.

Assuming separability, we model the covariance between the process at two different spatial locations and parameter settings $Y(\bs,\btheta)$ and $Y(\bs',\btheta')$ by 
\begin{equation*}
\Cov(Y(\bs,\btheta),Y(\bs',\btheta'))=K_s (\bs,\bs';\bxi_s) K_{\theta}(\btheta,\btheta';\bxi_\theta),
\end{equation*} 
where $K_s$ and $K_{\theta}$ are valid covariance functions respectively in $\mathcal{S}$ and $\boldsymbol{\Theta}$ with parameters $\bxi_s$ and $\bxi_\theta$. The covariance between discrepancy process $\bs$ and $\bs'$ is given by  
\begin{equation*}
\Cov(\delta(\bs),\delta(\bs'))=K_d (\bs,\bs';\bxi_d)
\end{equation*} 
with a valid covariance function $K_d$ in $\mathcal{S}$ and a vector of parameters $\bxi_d$. More specific definition of the covariance functions will be discussed below.

{\bf \noindent Computer Model Emulation. \label{section:emulation}}
The first component of our composite likelihood is the model for block
means, which captures the large scale trend. The covariance between the block means is
\begin{equation*}
\Sigma^{\barY}=\Eta \otimes \Sigma_\theta,
\end{equation*}
where $\Sigma_{\theta}$ is the covariance matrix for the random variable across $p$ parameter
settings and $\Eta$ is the $M\times M$ covariance matrix between the blocks.
It is straightforward to see that the block covariance is
\begin{equation} \label{eqn:computeH}
\{ \Eta\}_{ij}=\frac{1}{n_i n_j}\sum_{k=1}^{n_i} \sum_{l=1}^{n_j} K_s (\bs_{ik},\bs_{jl};\bxi_s),
\end{equation}
the mean of all possible cross covariances between two blocks.

The second component is the sum of the conditional likelihoods for
each block, which models the small scale dependence and variation.
For the $i$th block, the conditional distribution of output
 $\bY_{(i)}$ given the block mean $\barY_{(i)}$
 is a normal distribution with the mean and covariance given by
\begin{eqnarray*}
\mu_i^{\bY|\barY} &=&E(\bY_{(i)}|\bar{\bY}_{(i)})= (\gamma^{(i)} / \{
\Eta\}_{ii} \otimes I_p) \bar{\bY}_{(i)}\\
\Sigma_i^{\bY|\barY} &=&Var(\bY_{(i)}|\bar{\bY}_{(i)})= (\Gamma_i -
\gamma^{(i)} \left(\gamma^{(i)}\right)^T / \{ \Eta\}_{ii}) \otimes \Sigma_\theta
\end{eqnarray*}
where
\begin{eqnarray*}
\{\gamma^{(i)}\}_j &=& \sum_{k=1}^{n_i} K_s (\bs_{ij},\bs_{ik};\bxi_s) /n_i,~j=1,\dots,n_i-1,\\
\{\Gamma_i \}_{jk} &=& K_s (\bs_{ij},\bs_{ik};\bxi_s) ,~j=1,\dots,n_i-1,~k=1,\dots,n_i-1.
\end{eqnarray*}
Here, $\Gamma_i$ is the spatial covariance matrix for the $i$th block and $\gamma^{(i)}$ is the $(n_i-1)\times1$ covariance vector between the $i$th block mean and the $i$th block locations.
The log composite likelihood function for the model output is then
\begin{eqnarray*}
\cl(\bxi_s,\bxi_\theta)\propto&-&\frac{1}{2}
\left(\log\left|\Sigma^{\barY}\right|+\bar{\bY}^T \left(\Sigma^{\barY}\right)^{-1} \bar{\bY} \right)\\
&-&\frac{1}{2} \sum_{i=1}^{M}\left( \log\left|\Sigma_i^{\bY|\barY}\right| + \left(\bY_{(i)}-\mu_i^{\bY|\barY}\right)^T \left(\Sigma_i^{\bY|\barY}\right)^{-1}\left(\bY_{(i)}-\mu_i^{\bY|\barY}\right)\right),
\end{eqnarray*}
We construct the emulator by finding the MLE of $\bxi_\theta$ and
$\bxi_s$, denoted by $\hat{\bxi}_\theta$ and $\hat{\bxi}_s$. The
computational cost for a single likelihood evaluation is reduced
from $\frac{1}{3} n^3$ flops to $\sum_{i=1}^M \sum_{j=i}^M n_i n_j+\frac{1}{3}(M^3)+
\frac{1}{3}\left(\sum_{i=1}^{M}(n_i-1)^3 \right)$ flops, where the first term is the computational cost for finding $\mbox{H}$. This is a reduction from $6.86\times10^{10}$ flops to $5.92\times10^7$ flops in the climate model calibration example in Section  \ref{section:application}.
 
{\bf \noindent Computer Model Calibration.}\label{subsection:calibration} 
We formulate the composite likelihood for observational data in the
same manner as above. Let $\Omega$ be the $M\times M$ covariance between
the $M$ block means of the discrepancy $\bdelta$, defined in the same way as
$\Eta$ with a different set of
parameters $\bxi_d$. The conditional mean and covariance for the
block means of observational data $\barZ$ are
\begin{eqnarray*}
\mu^{\barZ}&=&\left( I_M \otimes \Sigma_{\theta^* \theta}
  \Sigma_\theta^{-1} \right) \barY,~\text{an}~M\times 1~\text{vector},\\
\Sigma^{\barZ} &=&
\Eta\otimes\left(\Sigma_{\theta^*}-\Sigma_{\theta^* \theta}
  \Sigma_\theta^{-1} \Sigma_{\theta^* \theta}^T  \right) +\Omega,~\text{an}~M\times M~\text{matrix}.
\end{eqnarray*}
Likewise, we define $\Lambda_i$ and $\lambda^{(i)}$ as the discrepancy
counterparts of $\Gamma_i$ and $\gamma^{(i)}$ with the covariance parameter
$\bxi_d$. Hence,  $\Lambda_i$ and $\lambda^{(i)}$ are the $i$th block discrepancy covariance matrix and the $(n_i-1)\times1$ covariance vector between the block outputs and the block mean respectively,
\begin{eqnarray*}
\{\lambda^{(i)}\}_j &=& \sum_{k=1}^{n_i} K_d(\bs_{ij},\bs_{ik};\bxi_d) /n_i,~j=1,\dots,n_i-1, \\
\{\Lambda_i \}_{jk} &=& K_d(\bs_{ij},\bs_{ik};\bxi_d),~j=1,\dots,n_i-1,~k=1,\dots,n_i-1.
\end{eqnarray*}
The conditional mean and covariance for observational data
in the $i$th block are therefore
\begin{eqnarray*}
\mu_{i}^{\bZ|\barZ}&=&\left( \mathbf{I}_{n_i-1} \otimes \Sigma_{\theta^* \theta}\Sigma_\theta^{-1} \right) \bY_{(i)}+(\tau^{(i)}+\lambda^{(i)})\left\{\Sigma^{\barZ} \right\}_{ii}^{-1} (\barZ_i - \left\{\mu^{\barZ} \right\}_i),\\
\Sigma_i^{\bZ|\barZ}&=&\left(\Gamma_i\otimes\left(\Sigma_{\theta^*}-\Sigma_{\theta^* \theta}
  \Sigma_\theta^{-1} \Sigma_{\theta^* \theta}^T  \right) +\Lambda_i\right)-(\tau^{(i)}+\lambda^{(i)})(\tau^{(i)}+\lambda^{(i)})^T /\left\{\Sigma_{\barZ}\right\}_{ii},
\end{eqnarray*}
where $\tau^{(i)} =\gamma^{(i)}\otimes \left(  \Sigma_{\theta^*} - \Sigma_{\theta^* \theta}\Sigma_{\theta}^{-1}\Sigma_{\theta^* \theta}^T \right)$. The log composite likelihood for the observational data is then
\begin{equation}
{\fontsize{10pt}{12pt}
\begin{split}
\cl_{n}(\bpsi)
\propto&-\frac{1}{2}\left(\log\left|\Sigma^{\barZ}\right|+\left(\barZ-\mu^{\barZ}\right)^T\left(\Sigma^{\barZ}\right)^{-1} \left(\barZ-\mu^{\barZ}\right)\right)\\
& -\frac{1}{2}\sum_{i=1}^{M} \left( \log\left|\Sigma_i^{\bZ|\barZ}\right| +\left( \bZ_{(i)} - \mu_i^{\bZ|\barZ} \right)^T \left( \Sigma_i^{\bZ|\barZ}\right)^{-1} \left( \bZ_{(i)} - \mu_i^{\bZ|\barZ} \right)\right),\label{eqn:CLobs}
\end{split}
}
\end{equation}
where the first line in \eqref{eqn:CLobs} is the log likelihood corresponding to the block means and the second line corresponding to the observations within each block.\ $\bpsi$ denotes all the parameters being estimated in the calibration stage including $\btheta^*$ and $\bxi_d$. By choosing a proper prior for $\bpsi$, $f(\bpsi)$, we define the approximate log posterior density, $\log(\pi_{n}(\bpsi)) \propto \log f(\bpsi)+\cl_{n}(\bpsi)$ and infer $\bpsi$ using the standard Metropolis-Hastings algorithm. We allow the scale parameters for the emulator to be re-estimated along with the other parameters but fix the other emulator parameters in $\bxi_s$ at their estimated values from the emulation stage \citep{bayarri2007computer, bhat:hara:tonk:kell:2009, chang2013fast}. The formulation  results in the same computation gain as in the emulation stage.

In both the emulation and calibration stages, calculation of the covariance matrix for the block means is a computational bottleneck, requiring $\sum_{i=1}^M \sum_{j=i}^M n_i n_j $ flops of computation. While computationally very demanding, its contribution to the likelihood function is usually not significant \citep{caragea2006approximate}. Therefore, instead of using all cross covariances between spatial locations, we randomly sample a subset of cross covariances to approximate the covariance between block means $\mbox{H}$. The computation of $\mbox{H}$ in \eqref{eqn:computeH} is substituted by
\begin{equation} \label{eqn:mean_approx}
\{ \Eta\}_{ij}=\frac{1}{m_i m_j}\sum_{k=1}^{m_i} \sum_{l=1}^{m_j} K_s(\bu_{ik},\bu_{jl};\bxi_s),
\end{equation}
with $m_i \le n_i$ and $m_j \le n_j$, where $\bu_{i1},\dots,\bu_{i m_i}$ and $\bu_{j1},\dots,\bu_{j m_j}$ are randomly chosen respectively from $\bs_{i1},\dots,\bs_{i n_i}$ and $\bs_{j1},\dots,\bs_{j n_j}$. This reduces the computational cost from $\sum_{i=1}^M \sum_{j=i}^M n_i n_j $ to $\sum_{i=1}^M \sum_{j=i}^M m_i m_j $ , that is, $1.32\times10^7$ flops to $2.86\times10^5$ flops for the calibration problem in Section \ref{section:application}. The same approximation can be applied to $\Omega$ with $\bxi_d$.
 
{\bf \noindent Covariance Function and Prior Specification.}
We use the exponential covariance function to define the covariance between parameter settings ($K_\theta$), spatial covariance for the emulator ($K_s$), and the spatial covariance for the discrepancy ($K_d$) with a nugget term. To be more specific, the covariance between the process at two parameter settings $\btheta=\left( \theta_1,\dots,\theta_q \right)^T$ and $\btheta'=\left( \theta'_1,\dots,\theta'_q \right)^T$ is defined by
\begin{equation*}
K_\theta(\btheta,\btheta';\bxi_\theta)=\zeta_\theta 1(\btheta=\btheta')+ \kappa_\theta \exp\left( -\sum_{i=1}^q \phi_{\theta,i} |\theta_i - \theta'_i |\right),
\end{equation*}
where $\bxi_\theta=\left(\zeta_\theta,\kappa_\theta, \phi_{\theta,1},\dots,\phi_{\theta,q} \right)$, and $\zeta_\theta, \kappa_\theta, \phi_{\theta,1},\dots,\phi_{\theta,q}>0$. Likewise,  the covariance between the process at two spatial locations $\bs$ and $\bs'$ for the emulator and the discrepancy term are given by
\begin{eqnarray*}
K_s(\bs,\bs';\bxi_s)=\kappa_s \left( \zeta_s 1(\bs=\bs')+  \exp\left( -\phi_s g(\bs , \bs')\right) \right),
\end{eqnarray*}
and
\begin{eqnarray}\label{equation:GPexpcov}
K_d(\bs,\bs';\bxi_d)= \kappa_d  \left( \zeta_d 1(\bs=\bs')+ \exp\left( -\phi_d g(\bs , \bs')\right) \right),
\end{eqnarray}
respectively, with $\bxi_s=\left( \zeta_s, \kappa_s, \phi_s \right)$, $\bxi_d=\left( \zeta_d, \kappa_d, \phi_d \right)$, and $\zeta_s, \kappa_s, \phi_s, \zeta_d, \kappa_d, \phi_d>0$.\ $g(\bs , \bs')$ denotes the distance between two points. In the climate model calibration problem in Section \ref{section:application}, for example, $g$ is the geodesic distance between two points on the earth's surface.

The parameters inferred by the Bayesian approach in the calibration stage are $\kappa_s$, $\zeta_d$, $\kappa_d$, $\phi_d$, and $\btheta^*$. Following \cite{bayarri2007computer}, the sill parameter for the emulator $\kappa_s$ is initially inferred via maximum likelihood estimate in the emulation stage and re-estimated by Bayesian inference in the calibration stage. We impose informative priors on the above parameters to avoid potentially obtaining improper posterior distributions \citep[cf.][]{berger2001objective} and identifiability issues. The latter is explained further in Section \ref{section:application}. The sill parameters, $\kappa_s$ and $\kappa_d$ receive inverse-Gamma priors $IG(a_{\kappa_s},b_{\kappa_s})$ and  $IG(a_{\kappa_d},b_{\kappa_d})$. We also impose an Inverse-Gamma prior $IG(a_{\zeta_d},b_{\zeta_d})$ for the nugget parameter $\zeta_d$. The prior density for the range parameter $\phi_d$ is assumed to be uniform with a wide support. The fitted computer model parameter $\btheta^*$ also receives a uniform prior over a wide range. Note that one can also assume a more informative prior for $\btheta^*$ such as a unimodal distribution based on some physical knowledge. However, in the calibration problem in Section \ref{section:application} we do not impose such a prior for $\btheta^*$; this allows us to study the characteristics of the posterior density of $\btheta^*$ more transparently.
 
{\bf \noindent Asymptotics and Adjustment using Godambe Information.}\label{subsection:adjustment}
Note that the composite likelihood in \eqref{eqn:CLobs} is not based
on the true probability model in \eqref{eqn:calibrationmodel}, and
therefore the `composite' posterior density based on \eqref{eqn:CLobs}
is quite different from the true posterior based on
\eqref{eqn:calibrationmodel}. In this section, we will discuss how the Godambe information matrix
\citep{godambe1960optimum} for estimating equations may be used to adjust for using the composite likelihood when making inferences.

We first provide the asymptotic justification for the adjustment using
the Godambe information matrix. We will show that, for large $n$ and $p$, the mode of the approximate posterior
$\hat{\bpsi}_{n}^B=\arg \max_{\bpsi} \pi_{n}(\bpsi)$ is consistent and
asymptotically normally distributed with a covariance matrix given by
the inverse of the Godambe information matrix. If we let $p \rightarrow \infty$, then the emulator converges to the measurement-error model such that
\begin{equation*}
\boldeta(\btheta) \sim N(\bY(\btheta), \zeta_\theta \Sigma^s),
\end{equation*}
where $\bY(\btheta)$ is the $n\times 1$ vector of  model output at the parameter setting $\btheta$ and the spatial locations $\bs_1,\dots,\bs_n$. This result holds as long as the computer model output varies reasonably smoothly in the parameter space \citep{yakowitz1985comparison}.
The model for observational data becomes
\begin{equation} \label{eqn:correct}
\bZ \sim N(\bY^*,\zeta_\theta \Sigma^s +\Sigma^d),
\end{equation} 
where $\bY^*=\bY(\btheta^*)$. The composite likelihood in \eqref{eqn:CLobs} then has the following means and covariances,
\begin{eqnarray*}
\mu^{\barZ}&=&\barY^*,~\text{an}~M\times 1~\text{vector},\\
\Sigma^{\barZ} &=&\zeta_\theta \Eta +\Omega,~\text{an}~M\times M~\text{matrix},\\
\mu_i^{\bZ|\barZ}&=&\bY_{(i)}^*+(\zeta_{\theta} \gamma^{(i)}+\lambda^{(i)})\left\{\Sigma^{\barZ} \right\}_{ii}^{-1} (\barZ_i - \left\{\mu^{\barZ} \right\}_i),\\
\Sigma_i^{\bZ|\barZ}&=&\left(\zeta_{\theta}\Gamma_i+\Lambda_i\right)-(\zeta_{\theta} \gamma^{(i)}+\lambda^{(i)})(\zeta_{\theta} \gamma^{(i)}+\lambda^{(i)})^T /\left\{\Sigma_{\barZ}\right\}_{ii},
\end{eqnarray*}
where $\barY_{(i)}^*=\frac{1}{n_i}\sum_{j=1}^{n_i} Y\left(\bs_{i
    	j},\btheta^*\right)$ is the $i$th block mean of the computer model
output at $\btheta^*$ and $\barY^*=\left(
  \barY_{(1)}^*,\dots,\barY_{(M)}^* \right)^T$ is the collection of
all their block means.

We now show the consistency and the asymptotic normality of the
posterior mode $\hat{\bpsi}_{n}^B$ as $n \rightarrow \infty$. We utilize expanding domain asymptotic results \citep[see e.g.][]{mardia1984maximum,cressie1993, cox2004note, zhang2005towards,varin2008composite}. The first step
is establishing consistency and asymptotic normality of the maximum
composite likelihood estimator. 
\begin{proposition}\label{proposition:regularity}
The following holds for the maximum composite likelihood estimator $\hat{\bpsi}_{n}^{CL}=\arg \max_{\bpsi} \cl_{n}(\bpsi)$;\\
(i) (Consistency) The maximum composite likelihood estimator is consistent for $\bpsi^0$;
\begin{equation*}
\hat{\bpsi}_{n}^{CL} \stackrel{\mathcal{P}}{\rightarrow} \bpsi^{0},
\end{equation*}
as $n \rightarrow \infty$, where $\bpsi^0$ is the vector of true values of parameters in $\bpsi$.\\
(ii) (Asymptotic Normality)  The asymptotic distribution of the maximum composite likelihood estimator is given by
\begin{equation*}
 \bG_n^{\frac{1}{2}} \left(\hat{\bpsi}_{n}^{CL}-\bpsi^0 \right)\stackrel{\mathcal{D}}{\rightarrow} N(0,\mathbf{I}) ,
\end{equation*}
 where $\bG_n=\bQ_n \bP_n^{-1} \bQ_n$ is the Godambe
information matrix \citep{godambe1960optimum}.\ $\bP_n$ is the
covariance matrix of the gradient $\triangledown \cl_{n}$ and $\bQ_n$ is the negative expected value of the Hessian matrix of $\cl_{n}$, where both are evaluated at $\bpsi=\bpsi^0$.
\begin{proof} For a composite likelihood, it is sufficient to verify the same regularity conditions as for the
  usual maximum likelihood estimators \citep{lindsay1988composite}. In the context of expanding domain asymptotics in spatial statistics, the spatial covariance function and its first and second derivatives need to be absolutely summable. From Theorem 3 in \cite{mardia1984maximum}, this condition holds for the exponential covariance function that we are using here. (i) and (ii) follow immediately.
\end{proof}
\end{proposition}
We are ready to state the main result of this section, which establishes the consistency and asymptotic normality of the posterior mode, $\hat{\bpsi}_n^B$.
\begin{proposition}\label{proposition:posterior}
(i) (Posterior consistency) The posterior degenerates on the true value $\bpsi^0$ in total variation, i.e.
\begin{equation}
|\pi_{n}(\bpsi)-\pi_n^0(\bpsi)|_{TV}\stackrel{\mathcal{P}}{\rightarrow}0
\label{eqn:consistency}
\end{equation}
as $n\rightarrow \infty$ where $|\cdot|_{TV}$ is the total variation norm and $\pi_n^0(\bpsi)$ is a normal density with the mean $\bpsi^0+\bQ_n^{-1}\triangledown \cl_{n}(\bpsi^0)$ and the covariance $\bQ_n^{-1}$. Note that $\bQ_n^{-1} \rightarrow \mathbf{0}$ as $n \rightarrow \infty$.\\
(ii) (Asymptotic normality) The density of $\hat{\bpsi}_{n}^B$ is asymptotically normal;
\begin{equation}
 \bG_n^{\frac{1}{2}} \left(\hat{\bpsi}_{n}^B-\bpsi^0 \right)\stackrel{\mathcal{D}}{\rightarrow} N(0,\mathbf{I}) ,
\label{eqn:Godambe}
\end{equation}
as $n \rightarrow \infty$.
\begin{proof}
When the maximum composite likelihood estimator $\bpsi_n^{CL}$ is consistent and asymptotically normal, (i) and (ii) follow \citep[Theorems 1 and 2 respectively in][]{chernozhukov2003mcmc}. Hence the result follows directly from Proposition \ref{proposition:regularity}.
\end{proof}
\end{proposition}
 
{\bf \noindent Application of Gobambe Adjustment.}\label{subsection:adjustment}
We have several options for adjusting our composite likelihood-based inference. These include 
(a) direct use of the asymptotic distribution in \eqref{eqn:Godambe}; (b) `open-faced sandwich' post-hoc adjustment \citep{shaby2012open} of MCMC sample from the composite posterior distribution $\pi_n(\bpsi)$; (c) `curvature' adjustment \citep{ribatet2009bayesian} for our MCMC procedure. We will utilize (b) and (c) because these MCMC-based methods can capture the higher-order moments of the posterior distribution, which may be important in finite sample inference.

For any of these methods, it is necessary to evaluate $\bP_n$ and $\bQ_n$. See the appendix for an example of their analytic computation. Note that $\bQ_n$ can also be obtained using MCMC runs from the posterior distribution $\pi_n(\bpsi)$ by the asymptotic result in  \eqref{eqn:consistency}.

We caution that the adjustment procedures here rely on the identifiability of parameters in $\bpsi$. In order to evaluate $\bP_n$ and $\bQ_n$ under the correct probability model in \eqref{eqn:correct}, we need to be able to estimate the true value $\bpsi^0$ accurately by the posterior mode $\hat{\bpsi}_n^B$. This may not always hold as there is a trade-off between the discrepancy parameters in $\bxi_d$ for finite sample sizes. 
 
The open-faced sandwich adjustment is one approach for adjusting the covariance based on Proposition \ref{proposition:posterior} \citep{shaby2012open}. For any MCMC sample of $\bpsi$ from $\pi_n(\bpsi)$, the open-faced sandwich adjustment is defined by $\tilde{\bpsi}^{open} =\hat{\bpsi}_n^B + \bC(\bpsi- \hat{\bpsi}_n^B)$ with $\bC=\bQ_n^{-1} \bP_n^{\frac{1}{2}} \bQ_n^{\frac{1}{2}}$. Similar to the curvature adjustment, this approach guarantees that the distribution of the adjusted posterior sample has the same posterior mode and the desired asymptotic covariance $\bG_n^{-1}$. Note that this method can be either embedded in each step of MCMC run or applied after an entire MCMC run is finished.

Another approach is curvature adjustment \citep{ribatet2009bayesian}, which substitutes $\bpsi$ in \eqref{eqn:CLobs} with $\tilde{\bpsi}^{curv} = \hat{\bpsi}_n^B + \bD(\bpsi - \hat{\bpsi}_n^B)$,
where $\bpsi$ is the posterior mode from \eqref{eqn:CLobs}.\ $\bD$ is the matrix that satisfies $\bD^T \bQ_n\bD = \bQ_n\bP_n^{-1} \bQ_n$. This approach ensures that the resulting posterior distribution has the same
mode as the original composite likelihood $\cl_n(\bpsi)$ and the
asymptotic covariance $\bG_n^{-1}$ as
described in \eqref{eqn:Godambe}. Note that the choice for $\bD$ is not unique, and \cite{ribatet2009bayesian} suggested using
$\bD=\bQ_n^{\frac{1}{2}}\left( \bQ_n\bP_n^{-1}\bQ_n\right)^{\frac{1}{2}}$ where the
square roots of the matrices are computed using singular value decomposition. Here we use the open-faced adjustment; the curvature adjustment approach may also be used but, as shown in  \citep{shaby2012open}, the difference between the two approaches is likely to be minimal.

\section{Application to UVic ESCM Calibration\label{section:application}}
{ We demonstrate the application of our approach to a climate model calibration problem. The computer model used here is the University of Victoria Earth system climate model (UVic ESCM) of intermediate complexity \citep{weaver2001uvic}. The input parameter that we are interested in is climate sensitivity (CS), defined as the equilibrium global mean surface air temperature change due to a doubling of carbon dioxide concentrations in the atmosphere \citep{andronova2007concept,knuttie2008equilibrium}. Climate sensitivity is an important model diagnostic and used as an input to climate projections as well as economic assessments of climate change impacts \citep[see e.g.][]{nordhaus2000warning,keller2004uncertain}. Each model run is a spatial pattern of ocean temperature anomaly on a regular $1.8^\circ$ latitude by $3.6^\circ$ longitude grid, defined as change between 1955-1964 mean and 2000-2009 mean in degree Celsius times meter ($^\circ$C m). At each location, the ocean temperature anomaly is vertically integrated from 0 to 2000 m. in depth.}

Note that the model output has regions of missing data since it covers only the ocean, and partition of the spatial area needs careful consideration. We partition the spatial area using a random tessellation; this is also the approach followed by\cite{eidsvik2011estimation}. We first randomly choose $M$ different centroids out of total $n$ locations and then assign the spatial locations to different subregions according to the nearest centroid in terms of geodesic distance. When finding the nearest centroid for each point, we only consider the centroids in the same ocean to avoid assigning locations separated by land to the same block. This random tessellation ensures, on average, that we have more subregions where data points are more densely distributed.

\subsection{Simulated Examples}
We conducted some perfect model experiments to answer the following questions: (i) Is the posterior density based on the composite likelihood (composite posterior) similar to the posterior density based on the original likelihood (original posterior)? (ii) Is the posterior density with approximated block mean covariance computation (approximated composite posterior) described in \eqref{eqn:mean_approx} close to the true composite  posterior? (iii) How do the number of spatial blocks and the magnitude of the discrepancy affect the composite posterior density?
 
Each experiment follows four key steps below:
\begin{enumerate}
\item Choose one of the parameter settings for model runs as the synthetic truth.
\item Leave the corresponding model run out and superimpose a randomly generated error on it to construct a synthetic observation.
\item Emulate the computer model using the remaining model runs.
\item Calibrate the computer model using the emulator in 3 and compare the resulting density with the synthetic truth.
\end{enumerate}
To be able to compute the original posterior density with a reasonable computational effort, we restrict ourselves to a subset of spatial locations consisting of 1000 randomly selected points and assume separable covariance structure for the spatial field and the computer model parameter space. The synthetic truth for the climate sensitivity used here is 2.153, but choosing other parameter settings gives similar results shown here.

A comparison between the composite posterior densities with 10 blocks and the original posterior densities are shown in Figure \ref{fig:comparison1} and \ref{fig:comparison2}. { We used two different realizations of the model-observation discrepancy. These were generated from a Gaussian process model with exponential covariance \eqref{equation:GPexpcov} with $\zeta_d^*=0.01$, $\kappa_d^*=160000$, and $\phi_d^*=690$ km, where $(\zeta_d^*, \kappa_d^*, \phi_d^*)$ are assumed true values of $(\zeta_d, \kappa_d, \phi_d)$.} We also conducted the same comparison for the approximated composite posterior densities (Figure \ref{fig:comparison_fast1} and \ref{fig:comparison_fast2}). The posterior densities and the resulting credible intervals from all three approaches are reasonably similar. The composite posterior densities after adjustment are slightly more dispersed than the original posterior due to the information loss caused by blocking, but the modes are quite close to the original ones confirming the consistency result in Proposition \ref{proposition:posterior} (i).

We also compared the adjusted composite posterior densities with different numbers of blocks to examine the effect of the number of blocks on calibration results (Figure \ref{fig:M}). The results show that using more than 30 blocks introduce a slight bias for the posterior mode which might be due to the reduced number of data points in each block. However, the credible intervals are again reasonably similar to each other. Similarly, we compare the adjusted composite posterior densities based on datasets generated using different assumed sill values, $\kappa_d^*=$40000, 90000 and 160000 to investigate the effect of magnitude of discrepancy on calibration results (Figure \ref{fig:kappa_d}). As one would expect, the posterior density becomes more dispersed as we increase the value of the sill.
 
We used informative priors for the statistical parameters, which is important to reduce the identifiability issues occurring in the calibration based on observational data in Section \ref{subsection:OBScalibration}. We imposed a vague prior for the nugget parameter $\zeta_d \sim IG(2,0.01(2+1))$ and a highly informative prior for the sill parameter $\kappa_d \sim IG(10000,\kappa_d^*(10000+1))$. The sill parameter for the emulator $\kappa_s$ is given a mildly informative prior with $IG(20,\hat{\kappa}_s (20+1))$, where $\hat{\kappa}_s$ is the MLE of $\kappa_s$ computed in the emulated stage. The shape parameters for the inverse-Gamma distributions are specified in the way that the prior modes are aligned with certain target values. Note that inference for simulated examples does not suffer from identifiability issues without the informative priors; we use these priors only to be consistent with the calibration based on observational data below.	
 
\subsection{Calibration using Observational Data} \label{subsection:OBScalibration}

{ As an illustrative example, we calibrate the climate sensitivity using the observed spatial pattern of ocean temperature anomaly from the data product constructed by \cite{levitus2012world}. We interpolated the observational data onto the UVic model grid using a simple bilinear interpolator. This step allows us to assume separability of emulation error and spatial covariance. We divide the 5,903 locations into 50 blocks using the random tessellation method described above. The covariance matrices for block means are approximated using \eqref{eqn:mean_approx} with $m_i=\mbox{min}(10,n_i)$ for $i=1,\dots,50$. The prior specification is the same as the simulated example with assumed sill ($\kappa_d^*$) of 160,000, except that the discrepancy range parameter $\phi_d$ is restricted to be greater than 800 km to reduce identifiability issues. Figure \ref{fig:obs} 
shows the posterior density of climate sensitivity. The length of the MCMC chain is 15,000, and the computing time is about 15 hours (wall time) via parallel computing using 32 high-performance cores for a system with Intel Xeon E5450 Quad-Core 434 at 3.0 GHz.} We verified that our MCMC algorithm and chain length were adequate by ensuring that the MCMC standard errors for our parameter estimates \citep{jones2006fixed, flegal2008markov} are small enough and by comparing posterior density estimates after various run lengths to see that the results, namely posterior pdfs, have stabilized.

\section{Discussion}
\subsection{Summary and future direction}
This work is, to our knowledge, the first application of composite likelihood to the computer model calibration problem. Our composite likelihood approach enables computationally efficient inference in computer model calibration using high-dimensional spatial data. We proved consistency and asymptotic normality of our posterior estimates and established covariance adjustment for posterior density based on them. The adjustment can be easily integrated into common MCMC algorithms such as the Metropolis-Hastings algorithm. The block composite likelihood used here yields a valid probability model, and therefore no additional verification for the propriety of the posterior distribution is necessary.

An attractive benefit of this general framework is that it is relatively easy, in principle, to extend the approach to a more complicated and easy-to-interpret covariance model. For example, by allowing covariance parameters to vary across the different spatial blocks, our approach can introduce non-stationarity in the spatial processes of model output and observational data. 

\subsection{Caveats}
While our approach is helpful in mitigating computational issues for various calibration problems, there is still more work to be done to make the computation more efficient. As $n$ continues to get large the number of spatial locations in each block may become excessively large and evaluation of composite likelihood may not be computationally tractable. One may consider increasing the number of blocks until the computation becomes feasible, but then the convergence of the posterior modes may be very slow due to too small block sizes \citep{cox2004note,varin2008composite}. Another perhaps simpler approach is to use a composite likelihood framework that does not involve blocks though this may involve the need for analytical work to establish posterior propriety.
 
Another possible issue is related to the use of a Gaussian emulator in place of the true computer model in computing $\bP_n$ and $\bQ_n$. Using a Gaussian process emulator, we approximate not only the true computer model itself, but also its first and second derivatives. In our particular example above, this does not cause any problem due to very regular behavior of the computer model output with respect to the input parameters. Note, however, that this may not be true in general and therefore $\bP_n$ and $\bQ_n$ calculations may be inaccurate.

It is also worth noting that the asymptotic independence between input parameters and discrepancy parameters does not usually hold in a finite sample. It is well known that calibration models usually suffer from identifiability issues \citep{wynn2001contribution}. One way to avoid the issues is imposing discrepancy prior information on the discrepancy term \citep{arendt2012quantification} as we did in Section \ref{section:application}.

{ The scientific result shown in \ref{subsection:OBScalibration} requires some caution in its interpretation. First, besides climate sensitivity, climate system response to changes in radiatively active gases in the atmosphere also depends on the magnitude of the radiative effects of these gases (``radiative forcing''), and on the vertical mixing of heat into the deep ocean \citep{hansen1985climate,knuttie2002constraints,schmittner2009using,urban2010probabilistic}. The parameters controlling both the forcing, and the vertical mixing, were kept fixed in the model runs we use. Including these additional uncertainties is expected to make the posterior density of CS more dispersed. 
The example serves as a demonstration of computational feasibility of our approach when applied to high-dimensional spatial datasets rather than providing an improved estimate of CS. Second, the variability of the posterior density is very sensitive to the prior information for the discrepancy term. Note, however, that this is a common problem for many calibration problems as discussed earlier.
\appendix
\section*{Appendix: Computation of $\bP_n$ and $\bQ_n$.}
In this supplementary material, we describe the matrix computation for $\bP_n=\Cov\left(\dot{\cl_n}(\bpsi)\right)$ and $\bQ_n=\E\left(\ddot{\cl_n}(\bpsi)\right)$. For ease of computation, it is useful to rewrite the composite likelihood function when $p=\infty$ in the following way:
\begin{equation*}
\begin{aligned}
\cl_{n}(\bpsi)
\propto&-\frac{1}{2}\left(\log|\Sigma^{\barZ}|+\left(\barZ-\barY^* \right)^T\left(\Sigma^{\barZ}\right)^{-1} \left(\barZ-\barY^*\right)\right)\\
 & -\frac{1}{2}\left( \sum_{i=1}^{M}\log|\Sigma_i^{\bZ|\barZ}| +\sum_{i=1}^M \left( \bZ_{[i]} - \bY_{[i]}^* \right)^T \bA_i^T\left( \Sigma_i^{\bZ|\barZ}\right)^{-1} \bA_i \left( \bZ_{[i]} - \bY_{[i]}^* \right)\right),
\end{aligned}
\end{equation*}
where $\bA_i$ is a $(n_i-1) \times n_i$ matrix such that
\begin{equation*}
\bA_i=\left( \mathbf{I}_{(n_i-1)\times (n_i-1)}~\mathbf{0}_{(n_i-1)\times 1} \right) - \bolda_i (\frac{1}{n_i} ,\dots, \frac{1}{n_i})_{1\times n_i},
\end{equation*}
and $\bolda_i$ is a $(n_i-1)\times 1$ vector such that
\begin{equation*}
\bolda_i=\left(\zeta_\theta \gamma^{(i)} + \lambda^{(i)} \right) \left\{ \Sigma^{\barZ} \right\}_{ii}^{-1}.
\end{equation*}
$\bZ_{[i]}$ is a $n_i \times 1$ vector containing all the $n_i$ observational data in the $i$th spatial block without omission, and $\bY_{[i]}$ is a $n_i \times 1$ vector of model output at $\btheta^*$ defined in the same way. Omitting the part irrelevant to the data, the partial derivative of $\cl_{n}(\bpsi) $ with respect to the $j$th computer model parameter, $\theta_j^*$, is given by
\begin{equation*}
\frac{\partial \cl_{n}(\bpsi) }{\partial \theta_j^*} \propto \barB_j^* (\barZ - \barY^*) + \sum_{i=1}^M \bB^*_{i,j} \left(\ \bZ_{[i]}-\bY_{[i]}^* \right),
\end{equation*}
where
\begin{eqnarray*}
\barB_j^*&=&\frac{\partial \barY^*}{\partial \theta_j^*} \left(\Sigma^{\barZ}\right)^{-1},\\
\bB^*_{i,j}&=&\left( \frac{\partial \bY_{[i]}^*}{\partial \theta_j^*} \right)^T \bA_i^T  \left( \Sigma_i^{\bZ|\barZ}\right)^{-1} \bA_i.
\end{eqnarray*}
We let $\bxi$ be the vector containing all the parameters in $\bxi_d$ as well as the emulator parameter being re-estimated.  The partial derivative with respect to the $k$th parameter in $\bxi$, $\xi_k$, can be written as
\begin{eqnarray*}
\frac{\partial \cl_{n}(\bpsi) }{\partial \xi_k} \propto && \frac{1}{2} (\barZ - \barY^*)^T \barB_k^d (\barZ - \barY^*) \\
&+&  \frac{1}{2} \sum_{i=1}^M \left(\ \bZ_{[i]}-\bY_{[i]}^* \right)^T \bB_{i,k}^d \left(\ \bZ_{[i]}-\bY_{[i]}^* \right) \\
&+& \sum_{i=1}^M \left(\ \bZ_{[i]}-\bY_{[i]}^* \right)^T \tilde{\bB}_{i,k}^d \left(\ \bZ_{[i]}-\bY_{[i]}^* \right)
\end{eqnarray*}
where
\begin{eqnarray*}
\barB^d_k&=& \left(\Sigma^{\barZ}\right)^{-1} \frac{\partial \Sigma^{\barZ}}{\partial \xi_k}  \left(\Sigma^{\barZ}\right)^{-1},\\
\bB_{i,k}^d&=& \bA_i^T \left(\Sigma^{\bZ|\barZ}_i\right)^{-1} \frac{\partial \Sigma^{\bZ|\barZ}_i}{\partial \xi_k}  \left(\Sigma^{\bZ|\barZ}_i\right)^{-1}  \bA_i,\\
\tilde{\bB}_{i,k}^d&=& - \left( \frac{\partial \bA_i}{\partial \xi_k} \right)^T  \left(\Sigma^{\bZ|\barZ}_i\right)^{-1} \bA_i.
\end{eqnarray*}
 
Note that inference on $\btheta^*$, our main goal, requires only calculating the asymptotic covariance of $\hat{\btheta}_n^B$ due to the asymptotic independence between $\hat{\btheta}_n^B$ and $\hat{\bxi}_n^B$, the posterior modes of $\btheta^*$ and $\bxi$ respectively. More specifically, for any $j$ and $k$,
\begin{eqnarray*}
\Cov\left(\frac{\partial \cl_{n}(\bpsi) }{\partial \theta_j^*} , \frac{\partial \cl_{n}(\bpsi) }{\partial \xi_k}\right)=0,
\end{eqnarray*}
because a linear combinations of zero-mean normal random variables and a quadratic form of the same variables are uncorrelated to one another. As a result, $\hat{\btheta}_n^B$ and $\hat{\bxi}_n^B$ have zero cross-covariance in $\bG_n$ and are asymptotically independent due to normality. Let $\bP_n^*$ be a part of $\bP_n$, which is the covariance matrix between partial derivatives with respect to the parameters in $\btheta^*$ only.  Likewise, let $\bQ_n^*$ be a part of $\bQ_n$ that contains only the negative expected Hessian of the parameters in $\btheta^*$. For inference on $\btheta^*$, it is sufficient to compute $\bP_n^*$ and $\bQ_n^*$ instead of $\bP_n$ and $\bQ_n$.
 
We compute the $(k,l)$th element of $\bP_n^*$ by plugging in $\hat{\bpsi}_n^B$ in place of $\bpsi$ in the following expression:
\begin{eqnarray*}
\Cov\left(\frac{\partial \cl_{n}(\bpsi) }{\partial \theta_k^*} , \frac{\partial \cl_{n}(\bpsi) }{\partial \theta_l^*}\right)=& &\barB_k^* \Sigma^{\barZ} \left(\barB_l^*\right)^T\\&+& \sum_{i=1}^M\sum_{j=1}^M \bB_{i,k}^*\ \Sigma_{i,j}^{\bZ} \left( \bB_{j,l}^* \right)^T\\ &+&\sum_{i=1}^M \barB_{k}^* \Sigma_{i}^{\barZ,\bZ} \left( \bB_{i,l}^* \right)^T\\
&+& \sum_{i=1}^M \barB_{l}^* \Sigma_{i}^{\barZ,\bZ} \left( \bB_{i,k}^* \right)^T,
\end{eqnarray*}
where $\Sigma_{i,j}^{\bZ}$ is the $n_i \times n_j$ covariance matrix between $\bZ_{[i]}$ and $\bZ_{[j]}$, and $\Sigma_{i}^{\barZ,\bZ}$ is the $1\times n_i$ covariance matrix between $\barZ$ and $\bZ_{[i]}$ under the probability model in (3). Similarly, the second order partial derivative of $\cl_{n}(\bpsi) $ with respect to $\theta_j^*$ and $\theta_k^*$ is given by
\begin{eqnarray*}
\frac{\partial \cl_{n}(\bpsi) }{\partial \theta_j^* \partial \theta_k^*} \propto && \left(\frac{\partial^2 \barY^*}{\partial \theta_j^* \partial \theta_k^*} \right)^T  \left(\Sigma^{\barZ}\right)^{-1} \left( \barZ-\barY^* \right)\\
&-&\left(\frac{\partial \barY^*}{\partial \theta_j^*}  \right)^T  \left(\Sigma^{\barZ}\right)^{-1} \frac{\partial \barY^*}{\partial \theta_k^*}\\
&+& \sum_{i=1}^M \left(\frac{\partial^2 \bY_{[i]}^*}{\partial \theta_j^* \partial \theta_k^*} \right)^T \bB_i^T  \left(\Sigma_i^{\bZ|\barZ}\right)^{-1} \bB_i \left( \bZ_{[i]}-\bY_{[i]}^* \right)\\
&-& \sum_{i=1}^M \left(\frac{\partial \bY_{[i]}^*}{\partial \theta_j^*} \right)^T \bB_i^T  \left(\Sigma_i^{\bZ|\barZ}\right)^{-1} \bB_i \frac{\partial \bY_{[i]}^*}{\partial \theta_k^*} .
\end{eqnarray*}
The $(j,k)$th element of $\bQ_n^*$ is computed by substituting $\bpsi$ with $\hat{\bpsi}_n^B$ in the following equation:
\begin{eqnarray*}
-E\left( \frac{\partial \cl_{n}(\bpsi) }{\partial \theta_j^* \partial \theta_k^*} \right) \propto & &\left(\frac{\partial \barY^*}{\partial \theta_j^*}  \right)^T  \left(\Sigma^{\barZ}\right)^{-1} \frac{\partial \barY^*}{\partial \theta_k^*}\\
&+& \sum_{i=1}^M \left(\frac{\partial \bY_{[i]}^*}{\partial \theta_j^*} \right)^T \bB_i^T  \left(\Sigma_i^{\bZ|\barZ}\right)^{-1} \bB_i \frac{\partial \bY_{[i]}^*}{\partial \theta_k^*} .
\end{eqnarray*}
 
Computing $\bP_n^*$ and $\bQ_n^*$ requires finding the first-order derivatives of $\bY_{[1]}^*,\dots,\bY_{[M]}^*$, and $\barY^*$. Since they are unknown functions of $\btheta^*$, we approximate them using the corresponding derivatives of the emulator output. The approximated derivatives of $\barY^*$ and $\bY_{[i]}^*$ with respect to $\theta_j^*$ are given by
\begin{eqnarray*}
\frac{\partial \barY^*}{\partial \theta_j^*}&=&\left( I_M \otimes
  \left( \frac{\partial \Sigma_{\theta^* \theta}}{\partial
    	\theta_j^*} \Sigma_{\theta}^{-1} \right) \right) \bar{\bY},\\
\frac{\partial \bY_{[i]}^*}{\partial \theta_j^* }&=&  \left(I_{n_i}\otimes \left(\frac{\partial \Sigma_{\theta^* \theta}}{\partial \theta_j^*} \Sigma_{\theta}^{-1}\right) \right)\bY_{[i]}.
\end{eqnarray*}
The derivative term $\frac{\partial \Sigma_{\theta^* \theta} }{\partial \theta_j^*}$ is determined by the covariance function for the parameter space. For the exponential covariance function used in our example, the derivative is
\begin{equation*}
\left\{ \frac{\partial \Sigma_{\theta^* \theta} }{\partial \theta_i^*} \right\}_j =\phi_{\theta,i}(-1)^{1(\theta_i^*>\theta_{ij})}\exp \left(-\sum_{k=1}^q \phi_{\theta,k}\left| \theta_k^*-\theta_{kj} \right| \right),~i=1,\dots,q,~j=1,\dots,p,
\end{equation*}
where $1(\cdot)$ is the indicator function, and $\theta_{ij}$ is the $i$th parameter value of the $j$th design point $\btheta_j$.

\bigskip
 
 
{\bf \noindent \large Acknowledgment}
This work was partly supported by NSF through the Network for Sustainable Climate Risk Management (SCRiM) under NSF cooperative agreement GEO-1240507 and the Penn State Center for Climate Risk Management (CLIMA). All views, errors, and opinions are solely that of the authors.
\clearpage

\newpage
\clearpage
 
\begin{figure}
\centering
\subfigure[realization 1 without approximation]{
\includegraphics[scale=0.35]{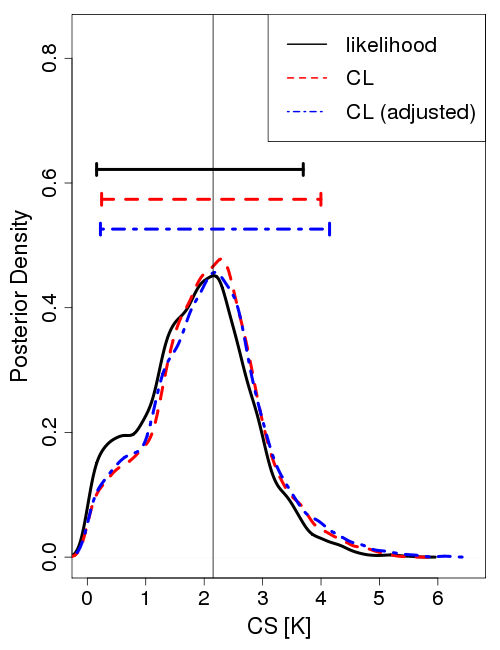}
\label{fig:comparison1}
}
\subfigure[realization 2 without approximation]{
\includegraphics[scale=0.35]{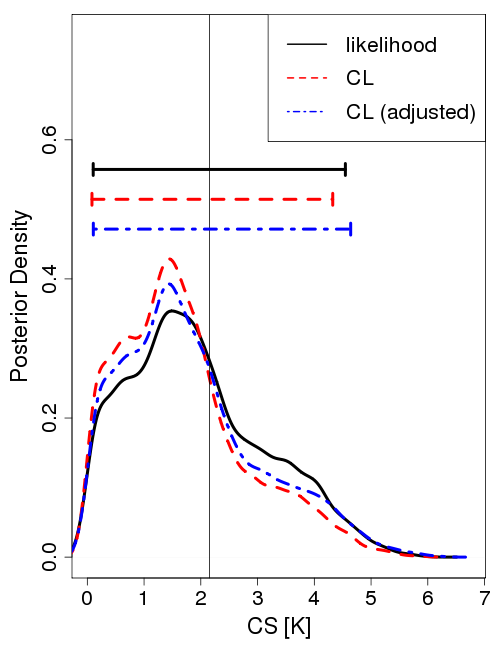}
\label{fig:comparison2}
}
\\
\subfigure[Realization 1 with approximation]{
\includegraphics[scale=0.35]{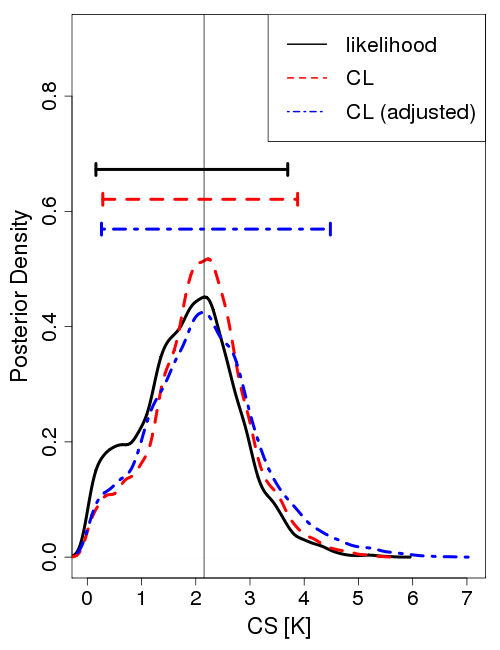}
\label{fig:comparison_fast1}
}
\subfigure[Realization 2 with approximation]{
\includegraphics[scale=0.35]{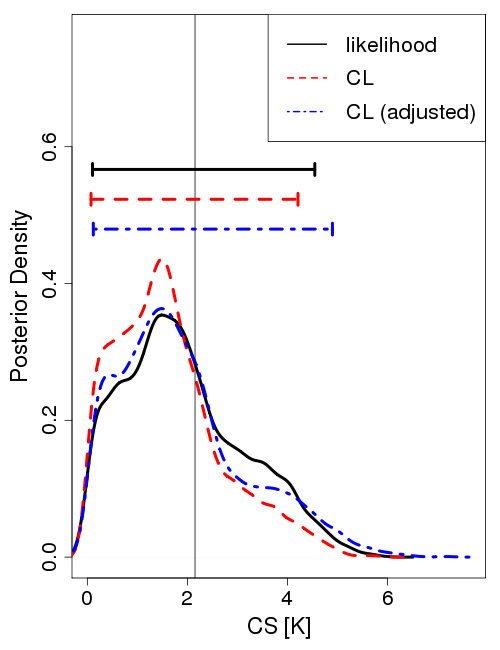}
\label{fig:comparison_fast2}
}
\caption{Comparison between calibration results using i) the original likelihood without blocking (solid black curves), ii) the block composite likelihood without the variance adjustment (dashed red line), and iii) the block composite likelihood with the variance adjustment (dashed-dotted blue line). The vertical lines represent the assumed true value for our simulation, and the horizontal bars above show the 95\% credible intervals. The results shown here are based on two different realizations (two left panels for Realization 1 and two right panels for Realization 2) from the same GP model. 
The posterior densities with the approximation for the block means (two lower panels) are reasonably close to the densities without the approximation (two upper panels) when the variance adjustment is applied. 
}
\label{fig:comparison}
\end{figure}
\begin{figure}
\centering
\includegraphics[scale=0.5]{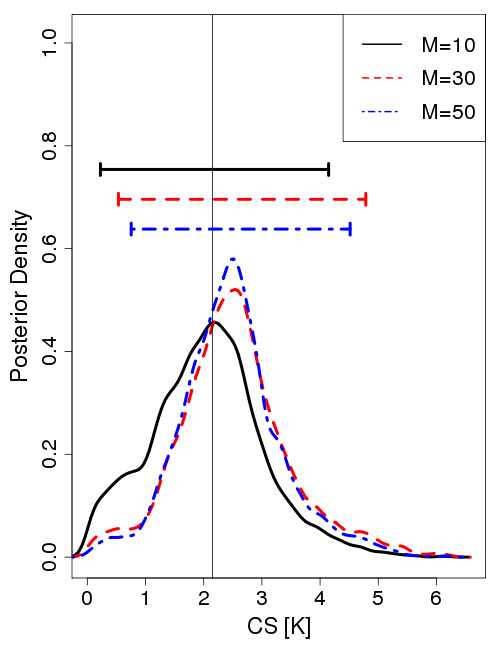}
\caption{Comparison of posterior densities between three simulated examples with different block numbers: $M=10$ (solid black curve), $M=30$ (dashed red curve), and $M=50$ (dotted-dashed blue curve). The vertical line is the assumed true value for our simulated example and the horizontal bars above are 95\% credible intervals. Posterior modes based on 30 and 50 blocks show slight biases, but the width of interval does not show notable differences.}.
\label{fig:M}
\end{figure}
\begin{figure}
\centering
\includegraphics[scale=0.5]{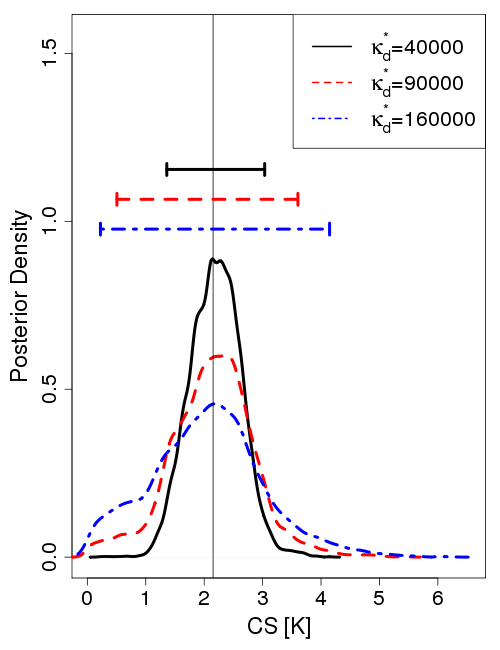}
\caption{Comparison of posterior densities between three simulated examples with different assumed magnitudes of the discrepancies: $\kappa_d^*=40000$ (solid black curve), $\kappa_d^*=90000$ (dashed red curve), and $\kappa_d^*=160000$ (dotted-dashed blue curve). The vertical line indicates the assumed true values, and the horizontal bars above show the 95\% credible intervals. As the discrepancy grows, the densities become more dispersed but the posterior modes stay similar.}
\label{fig:kappa_d}
\end{figure}
\begin{figure}
\centering
\includegraphics[scale=0.5]{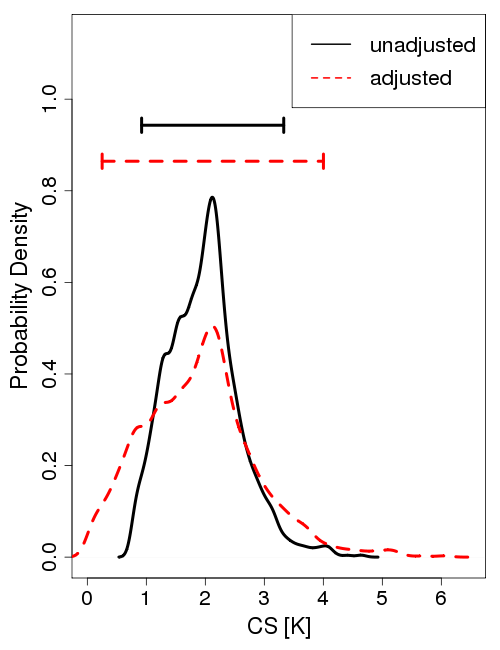}
\caption{Posterior densities of the climate sensitivity calibrated based on the observational data from \cite{levitus2012world} using our composite likelihood approach. The adjusted posterior density (solid black curve) is notably more dispersed than the unadjusted one (dashed black curve), and the corresponding 95\% credible intervals (horizontal bars above) for the adjusted posterior density is also much wider than the one for the unadjusted density.}
\label{fig:obs}
\end{figure}

\clearpage
 
\bibliographystyle{asa_original}
\bibliography{short,references}
\end{document}